\documentclass[reprint,superscriptaddress,
 amsmath,amssymb,
 aps,
prl,
]{revtex4-2}

\usepackage{graphicx}
\usepackage{dcolumn}
\usepackage{bm}

\usepackage{xcolor}
\usepackage{float}
\usepackage{booktabs}
\usepackage{siunitx}
\begin{document}
\title{A Transport Framework for Evaluating Nanoscale Interconnect Materials}
\author{Md. Rafiqul Islam}
\email{Corresponding Author: mrisagor013@gmail.com}
\affiliation{Department of Mechanical and Aerospace Engineering, University of Virginia, Charlottesville, Virginia 22904, USA}
\author{Patrick E. Hopkins}%
   \email{Corresponding Author: phopkins@virginia.edu}
  \affiliation {%
Department of Mechanical and Aerospace Engineering, University of Virginia, Charlottesville, Virginia 22904, USA}%
  \affiliation{%
Department of Materials Science and Engineering, University of Virginia, Charlottesville, Virginia 22904, USA}%
 \affiliation{%
Department of Physics, University of Virginia, Charlottesville, Virginia 22904, USA}%

\begin{abstract}
    As CMOS technology continues to scale, metallic interconnects increasingly limit circuit performance through rising resistivity, self-heating, and reliability degradation. Although several alternative metals have been proposed, a quantitative framework for evaluating their transport performance under nanoscale confinement remains unavailable. Here, we establish a transport framework by combining independently measured thermal and electrical conductivities with normalized and unnormalized transport figures of merit. We apply this framework to Cu, Ru, W, Co, Ir, and Mo thin films using new steady-state thermoreflectance measurements of Mo, Co, and Ir together with previously reported Cu, Ru, and W data. While Cu exhibits the highest intrinsic transport performance, its effective performance is substantially reduced by thickness scaling and Ta liner resistance. In contrast, Ru and Mo maintain favorable transport properties while enabling barrierless integration, identifying them as promising candidates for next-generation CMOS interconnects.
\end{abstract}

\keywords{thin films, Matthiessen's rule, thermoreflectance, in-plane thermal conductivity, electron-phonon interactions}

\maketitle

Moore’s law continues to drive transistor scaling into the nanometer regime, resulting in elevated power densities and severe localized heating~\cite{krishnan2007towards,seshan2012scaling}. In this limit, interconnects have emerged as a primary performance bottleneck because their electrical and thermal transport properties deviate significantly from bulk values. Specifically, increased resistivity and self-heating in scaled interconnects degrade the reliability and efficiency of advanced integrated circuits~\cite{zhan2020effect}. Copper (Cu), the industry-standard material, exhibits a precipitous rise in resistivity near 25~nm due to enhanced surface and grain boundary scattering~\cite{islam2024evaluating}. While recent evaluations have integrated sustainability metrics, such as supply risk and environmental impact, into the selection criteria~\cite{soulie2024selecting, boakes2024selection}, resistivity remains the critical metric, as it fundamentally dictates signal delay and power dissipation~\cite{perez2022dominant, gall2020search, islam2024evaluating}.

\begin{figure}[h!]
\centering \includegraphics[scale=0.6]{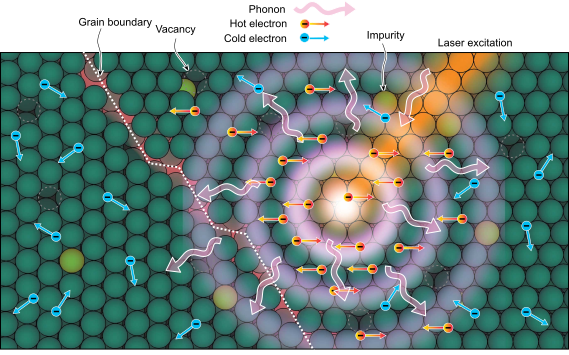} 
\caption{Microscopic scattering processes in laser-excited metallic thin films. The diagram depicts the interplay between excited electrons and the lattice (phonons), as well as structural defects including grain boundaries and atomic vacancies. These interactions define the characteristic length scales that limit charge and heat transport as device dimensions scale toward the nanometer regime } 
\label{Fig01}
\end{figure}

Electromigration presents a parallel reliability challenge. As cross-sectional areas vanish, increased current densities accelerate atomic transport, leading to catastrophic void formation and hillock growth~\cite{black1969electromigration}. While Cu replaced aluminum for its superior resistance~\cite{edelstein1997full}, scaled Cu lines now suffer from enhanced grain boundary diffusion and the parasitic volume of necessary liner/barrier layers~\cite{zahedmanesh2019copper}.As VLSI nodes scale below 5~nm, the search for alternative materials has intensified~\cite{zhan2020effect}. These nanoscale challenges are rooted in the fundamental physics governing heat and charge transport. In bulk metals, electron-driven heat transfer is predominantly limited by electron-phonon ($e$-$ph$) interactions at room temperature. However, as structures shrink below the intrinsic $e$-$ph$ mean free path, boundary scattering emerges as the dominant mechanism, significantly reducing transport efficiency. At these scales, conductivities are deleteriously influenced by complex interactions between excited electrons, phonons, grain boundaries, and defects, as conceptually illustrated in Fig.~\ref{Fig01}. Promising candidates such as ruthenium (Ru), cobalt (Co), molybdenum (Mo), and iridium (Ir) offer shorter electron mean free paths~\cite{gall2016electron}, resulting in a less dramatic resistivity scaling compared to Cu. Furthermore, relatively high cohesive energies and melting points of these aforementioned metals suggest intrinsically superior electromigration resistance~\cite{haynes2016crc, soulie2024selecting} and the potential for barrierless integration. This directly links fundamental bonding energy and scattering cross-sections to the long-term reliability of next-generation devices.

\begin{figure*}[t]
\centering
\includegraphics[width=\textwidth]{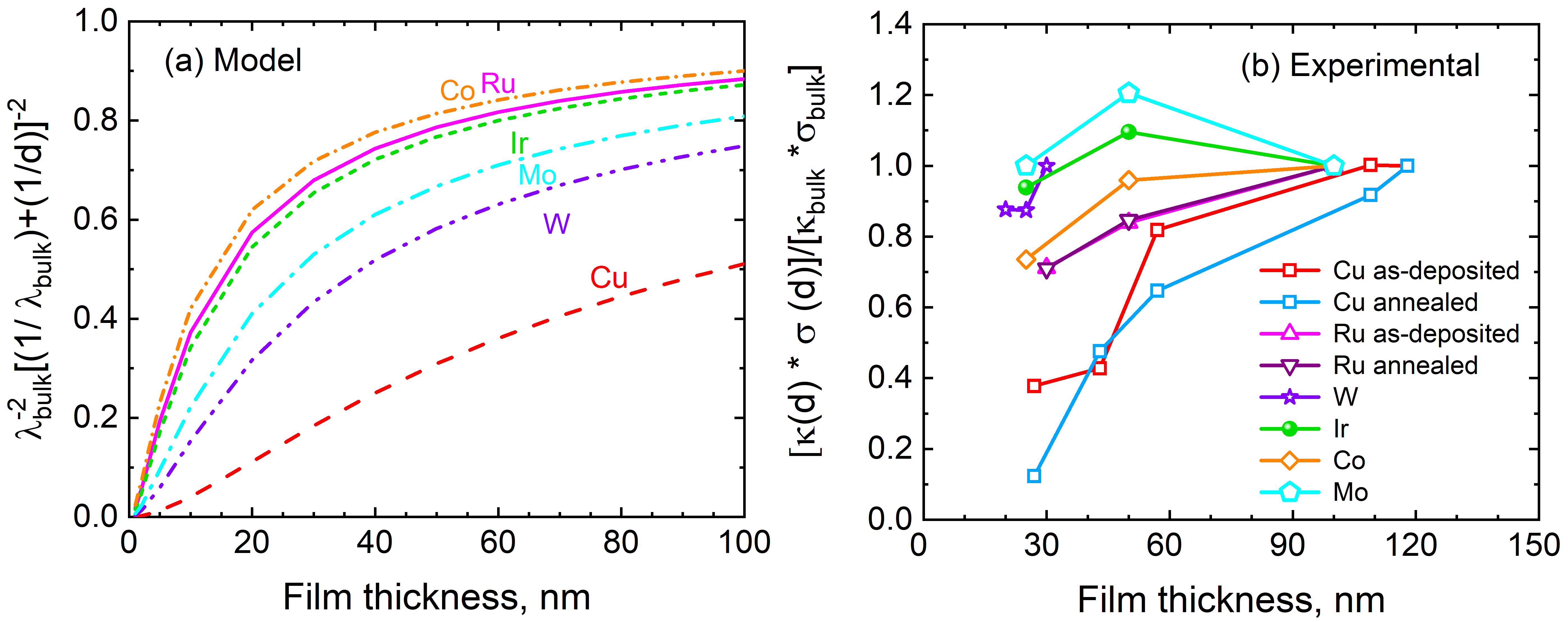}
\caption{(a) Theoretical predictions for candidate metals incorporating mean-free-path limitations, highlighting the stronger thickness-dependent degradation of Cu relative to the alternative metals. (b) Experimental comparison of the combined thermal and electrical transport performance of Cu, Ru, W, Ir, Co, and Mo thin films as a function of thickness. The plotted data correspond to the nominal thermal and electrical conductivities used in the analysis.}
\label{Fig02}
\end{figure*}
In this work, we employ a scaling-based analytical framework to evaluate the coupled thermal and electrical transport behavior of metallic interconnects under nanometric confinement. By integrating the Wiedemann--Franz relationship with a thickness-dependent mean free path model, we establish a predictive figure of merit (FOM) that quantifies how intrinsic scattering mechanisms limit the overall transport performance of candidate interconnect metals. We apply this framework to Ir, Mo, Co, Cu, Ru, and W using both intrinsic ($\kappa\sigma$) and effective ($\kappa_{\mathrm{eff}}\sigma_{\mathrm{eff}}$) transport products to assess their viability for nanoscale interconnect applications. To validate the proposed framework, the predicted FOM is compared against experimentally measured thermal conductivities of PVD-grown metallic thin films with thicknesses ranging from $25$ to $100~\mathrm{nm}$ using steady-state thermoreflectance (SSTR). 

Our results identify Ru and Mo as particularly promising candidates, exhibiting reduced resistivity scaling and enhanced transport efficiency relative to the other metals investigated. By combining independent thermal and electrical transport measurements, this work establishes a comprehensive framework for evaluating nanoscale interconnect materials and provides a critical experimental dataset for the selection of metals in next-generation integrated circuits.

To compare candidate interconnect metals, we define a transport figure of merit (FOM) that combines thermal and electrical transport while accounting for thickness-dependent boundary scattering. Assuming (i) the Wiedemann–Franz relation for the electronic contribution to thermal conductivity, (ii) Matthiessen's rule for the effective electron mean free path via Kinetic Theory considerations, and (iii) thermal and electrical conductivities scale with the effective mean free path, the normalized FOM becomes

\begin{equation}
\text{FOM} = \lambda_{\text{bulk}}^{-2}
\left(
\frac{1}{\lambda_{\text{bulk}}}
+
\frac{1}{d}
\right)^{-2}
\end{equation}

where \begin{equation}
    \lambda(d) = \left( \frac{1}{\lambda_{\text{bulk}}} + \frac{1}{d} \right)^{-1}
\end{equation}

Here, $\lambda(d)$ is the effective electron mean free path in a film of thickness $d$, while $\lambda_{\mathrm{bulk}}$ denotes the intrinsic bulk electron mean free path. As the film thickness decreases, boundary scattering shortens the effective mean free path, resulting in reduced thermal and electrical transport. The complete derivation and  further discussions of  the aforementioned assumptions are provided in the Supporting Information.

Figure~\ref{Fig02} compares the transport performance of candidate interconnect metals as the film thickness decreases. Figure~\ref{Fig02}(a) presents theoretical predictions based on the scattering-limited scaling model and the Wiedemann--Franz law, assuming both $\kappa$ and $\sigma$ scale with the effective electron mean free path. Bulk electron mean free paths are taken from Ref.~\cite{gall2016electron}. The model predicts that Cu exhibits the most pronounced degradation owing to its relatively long bulk mean free path ($\lambda_{\mathrm{bulk}}\approx39.9$~nm). In contrast, metals with shorter mean free paths, including Ru ($\approx6$~nm), Co ($\approx11.8$~nm), and Ir ($\approx7.1$~nm), retain their transport performance even below 10~nm, consistent with the electrical resistivity analysis of Gall~\cite{gall2020search}. The model here assumes that the Wiedemann--Franz law applies to the electronic contribution to thermal conductivity and neglects interface and adhesion-layer effects. Since the Wiedemann--Franz law may not accurately describe the total thermal conductivity when phonons contribute to heat transport, we independently measure the thermal conductivity ($\kappa$) and electrical conductivity ($\sigma$) using steady-state thermoreflectance (SSTR) and four-point probe measurements, respectively. These independent measurements enable both validation of the proposed transport figure of merit and assessment of the applicability of the Wiedemann--Franz relation for the investigated metals.

The in-plane thermal conductivities of the metallic thin films are measured using the steady-state thermoreflectance (SSTR) technique, which we describe in detail elsewhere~\cite{braun2019steady,islam2024evaluating,islam2026unveiling}. Briefly, SSTR is a low-frequency pump--probe technique that establishes steady-state thermal gradients, providing high sensitivity to the in-plane thermal conductivity of thin films. To ensure consistent optothermal transduction and suppress substrate-induced optical artifacts, we coat all samples with a 20~nm Al/60~nm Ti transducer. This transducer provides optical opacity at the probe wavelength while minimizing lateral heat spreading, thereby maximizing sensitivity to the in-plane thermal conductivity of the underlying metal film. 

\begin{figure*}[!ht]
\centering
\includegraphics[width=\textwidth]{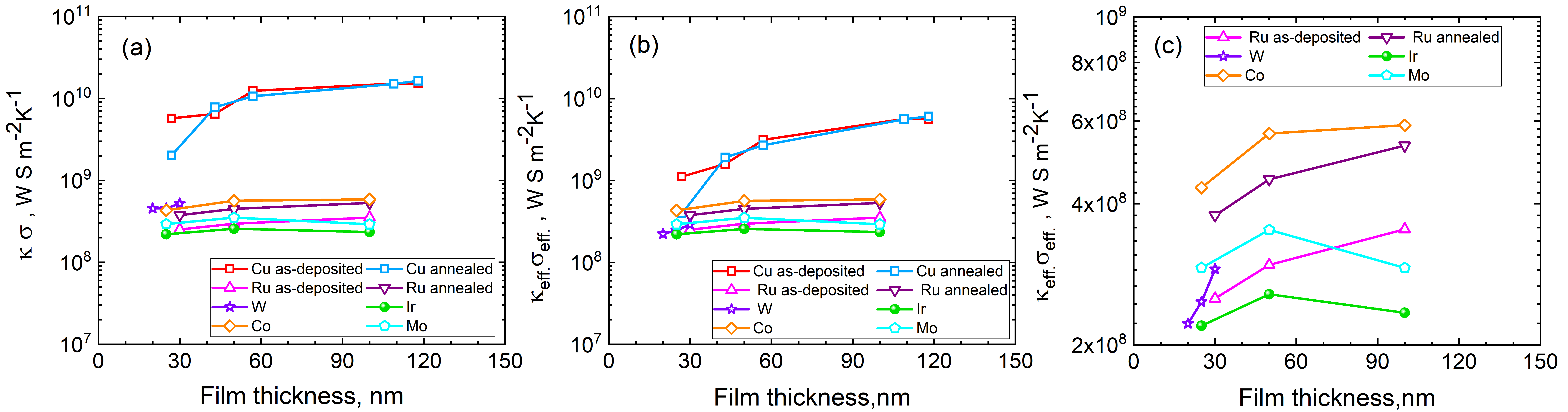}
\caption{
(a) Measured $\kappa\sigma$ values for Cu, Ru, W, Co, Mo, and Ir thin films. Cu shows the highest $\kappa\sigma$ for thicknesses above 20~nm. (b) The impact of a 3~nm Ta adhesion layer is considered for Cu, leading to a reduction in its effective $\kappa\sigma$ and eliminating its performance advantage at lower thicknesses. (c) Effective $\kappa_{\mathrm{eff}}\sigma_{\mathrm{eff}}$ values for barrierless Ru, W, Co, Mo, and Ir thin films, demonstrating their transport performance in the absence of adhesion and diffusion-barrier layers. The data in the figure represent the nominal thermal and electrical conductivities used in the transport analysis.}
\label{fig03}
\end{figure*}
We measure the electrical conductivity of each film using a standard four-point probe technique and estimate the electronic thermal conductivity from the Wiedemann--Franz relation using the Sommerfeld value of the Lorenz number, $L_0 = 2.44 \times 10^{-8}~\mathrm{W\,\Omega\,K^{-2}}$. Although the Lorenz number may deviate from the Sommerfeld value in thin metallic films~\cite{islam2026unveiling}, this estimate provides a useful reference for comparison with the independently measured thermal conductivity, thus allowing us to assess the validity of this widely implemented assumption in Wiedemann-Franz law-based thermal conductivity calculations.

The Cu, Ru, and W datasets are obtained from our previous studies~\cite{islam2024evaluating,islam2026unveiling}, whereas we measure the Mo, Co, and Ir films as part of the present study. We provide additional details on sample preparation, measurement techniques, data analysis, and uncertainty calcualation  in the Supporting Information. The experimental results qualitatively agree with the model predictions and support the selection of alternative interconnect metals beyond Cu at nanoscale dimensions, as shown in Fig.~\ref{Fig02}(b). Cu exhibits the most pronounced reduction in transport performance with decreasing thickness, and annealing further degrades its performance because of defect formation~\cite{islam2024evaluating}. In contrast, Ru, W, Co, and Ir exhibit substantially weaker thickness dependence, consistent with their shorter electron mean free paths. Among these materials, Mo and Ir retain the highest transport performance over the investigated thickness range.

Normalized figures of merit facilitate comparisons relative to bulk properties but can obscure differences in the absolute transport performance relevant to practical interconnect applications. We therefore evaluate both the normalized and unnormalized $\kappa\sigma$ values. The unnormalized $\kappa\sigma$ directly reflects the measured thermal and electrical transport of each film, whereas the normalized metric isolates the effect of thickness-dependent transport degradation. For example, annealing produces only minor changes in the normalized performance of Cu and Ru, while the unnormalized $\kappa\sigma$ reveals an improvement in Ru arising from microstructural modifications, as discussed below. Unless otherwise noted, the analysis excludes the thermal and electrical contributions of the 3~nm Ta adhesion layer.

Figure~\ref{fig03}(a) presents the unnormalized $\kappa\sigma$ values for all metals based on the experimentally measured values from Fig.~\ref{Fig02}b. Cu exhibits the highest transport performance for thicknesses exceeding approximately 20~nm because of its intrinsically high electrical and thermal conductivity. In contrast, W and Ir show lower $\kappa\sigma$ values over the investigated thickness range, with Ir exhibiting the lowest transport performance owing to its relatively low electrical conductivity. Although normalized metrics emphasize transport degradation relative to bulk behavior, the unnormalized $\kappa\sigma$ provides a more direct comparison of the transport performance expected in practical interconnects.

The practical implementation of Cu interconnects requires adhesion and diffusion-barrier layers to suppress Cu diffusion into the surrounding dielectric, thereby preventing dielectric breakdown and interconnect failure~\cite{lo2019enhancing,zhan2020effect}. To account for this practical constraint, Fig.~\ref{fig03}(b) incorporates a 3~nm Ta adhesion/barrier layer into the transport analysis. The electrical conductivity of Ta is obtained from its measured electrical resistivity, while its electronic thermal conductivity is estimated using the Wiedemann--Franz relation with the Sommerfeld Lorenz number. The effective electrical and thermal transport properties of the Cu/Ta stack are then calculated by treating the Cu conductor and Ta liner as electrical and thermal resistances connected in series, respectively. Owing to the relatively low electrical and thermal conductivity of Ta, its inclusion substantially degrades the effective transport performance of Cu, particularly at reduced dimensions where the liner occupies an increasingly large fraction of the interconnect cross section. Details of the transport model are provided in the Supporting Information.

Among the candidate materials, Ru, Mo and Cu provide a favorable balance between transport performance and practical integration as shown in Fig.~\ref{fig03}c. Unlike Cu, both materials are compatible with barrierless integration while exhibiting excellent electromigration resistance~\cite{soulie2024selecting,haynes2016crc}. Previous studies further show that Mo adheres well to dielectric materials, including SiO$_2$, low-$\kappa$ organosilicate glass, and SiCO, without requiring adhesion or barrier layers~\cite{lesniewska2020dielectric,lesniewska2021reliability}. Although Co exhibits competitive transport performance (Fig.~\ref{fig03}c) , its adhesion reliability and resistance to delamination require further investigation~\cite{griggio2018reliability}.

By combining normalized and unnormalized transport figures of merit with independently measured thermal and electrical conductivities, we establish a quantitative framework for evaluating nanoscale interconnect materials. The results demonstrate that intrinsic transport performance alone is insufficient for material selection once thickness scaling and integration constraints are considered. Within this framework, Ru and Mo emerge as the most promising candidates for future logic and memory interconnects.

\section*{Acknowledgements}
This work was supported as part of APEX (A Center for Power Electronics Materials and Manufacturing Exploration), an Energy Frontier Research Center funded by the U.S. Department of Energy, Office of Science, Basic Energy Sciences under Award No ERW0345. We thank Sean W.~King, Colin D.~Landon, Christopher Jezewski, Rinus T.~P.~Lee, and Kandabara Tapily for providing the thin-film samples. We also acknowledge Intel Corporation and TEL Technology Center, America, LLC, for their support in thin-film synthesis.

\clearpage
\onecolumngrid
\setcounter{section}{0}
\renewcommand{\thesection}{S\arabic{section}}
\setcounter{figure}{0}
\renewcommand{\thefigure}{S\arabic{figure}}
\setcounter{table}{0}
\renewcommand{\thetable}{S\arabic{table}}
\setcounter{equation}{0}
\renewcommand{\theequation}{S\arabic{equation}}

\begin{center}
{\Large\bfseries Supporting Information}\par
\vspace{0.8em}
{\large\bfseries A Transport Framework for Evaluating Nanoscale Interconnect Materials}\par
\vspace{0.8em}
Md. Rafiqul Islam and Patrick E. Hopkins\par
\vspace{0.4em}
Department of Mechanical and Aerospace Engineering, University of Virginia, Charlottesville, Virginia 22904, USA\par
\end{center}
\vspace{1em}
\section{Synthesis and Sample Description}

We synthesize Cu, Ru, W, Mo, Co, and Ir thin films with thicknesses ranging from ~5 to 100~nm by physical vapor deposition (PVD). Cu and W films are deposited on 3~nm Ta/100~nm SiO$_2$/Si substrates, whereas Ru, Mo, Co, and Ir films are deposited directly on 100~nm SiO$_2$/Si substrates without an adhesion or barrier layer. The thermal transport properties of the Cu, Ru, and W films have been reported in our previous studies~\cite{islam2024evaluating,islam2026unveiling}, whereas the Ir, Mo, and Co films are characterized in the present work. The Ir, Mo and Co films are deposited by Intel on SiO$_2$/Si substrates.

\section{Derivation of the Transport Figure of Merit}

As interconnect dimensions continue to shrink in advanced microelectronics, the transport performance of metallic thin films becomes increasingly sensitive to surface and grain-boundary scattering. An ideal interconnect material maintains high thermal conductivity, low electrical resistivity, and minimal transport degradation as the film thickness decreases. The thickness dependence of thermal and electrical transport is commonly characterized by normalizing the measured conductivities to their corresponding bulk values~\cite{chavez2014reduction,fuchs1938conductivity},

\begin{equation}
\frac{\kappa(d)}{\kappa_{\mathrm{bulk}}}, \qquad
\frac{\sigma(d)}{\sigma_{\mathrm{bulk}}},
\end{equation}

where $\kappa(d)$ and $\sigma(d)$ denote the thermal and electrical conductivities of a film with thickness $d$, and $\kappa_{\mathrm{bulk}}$ and $\sigma_{\mathrm{bulk}}$ are the corresponding bulk values. Assuming the Wiedemann--Franz (WF) law is applicable to the electronic contribution to thermal conductivity,

\begin{equation}
\kappa=L_0\sigma T,
\end{equation}

where $L_0$ is the Sommerfeld Lorenz number and $T$ is the absolute temperature. Combining the thermal and electrical transport properties gives

\begin{equation}
\kappa\sigma=\frac{\kappa^2}{L_0T},
\end{equation}

which serves as a measure of the coupled thermal and electrical transport capability of the metal. To account for finite-size effects, we describe the effective electron mean free path using Matthiessen's rule,

\begin{equation}
\lambda(d)=
\left(
\frac{1}{\lambda_{\mathrm{bulk}}}
+\frac{1}{d}
\right)^{-1},
\end{equation}

where $\lambda(d)$ is the effective electron mean free path of a film with thickness $d$, and $\lambda_{\mathrm{bulk}}$ denotes the intrinsic bulk electron mean free path. As the film thickness decreases, enhanced boundary scattering reduces the effective mean free path and consequently suppresses both thermal and electrical transport. Within the Fuchs--Sondheimer size-effect framework, both the electrical conductivity and the electronic contribution to the thermal conductivity scale approximately with the effective electron mean free path,

\begin{equation}
\kappa(d)\propto\lambda(d), \qquad
\sigma(d)\propto\lambda(d).
\end{equation}

Therefore,

\begin{equation}
\kappa(d)\sigma(d)
\propto
\lambda(d)^2
=
\left(
\frac{1}{\lambda_{\mathrm{bulk}}}
+\frac{1}{d}
\right)^{-2}.
\end{equation}

We define the normalized transport figure of merit (FOM) as

\begin{equation}
\mathrm{FOM}
=
\frac{\kappa(d)\sigma(d)}
{\kappa_{\mathrm{bulk}}\sigma_{\mathrm{bulk}}}.
\end{equation}

Since the bulk transport product scales with the square of the bulk electron mean free path,

\begin{equation}
\kappa_{\mathrm{bulk}}\sigma_{\mathrm{bulk}}
\propto
\lambda_{\mathrm{bulk}}^2,
\end{equation}

the normalized transport figure of merit becomes

\begin{equation}
\boxed{
\mathrm{FOM}
=
\lambda_{\mathrm{bulk}}^{-2}
\left(
\frac{1}{\lambda_{\mathrm{bulk}}}
+\frac{1}{d}
\right)^{-2}
}
\end{equation}

This expression describes the degradation of coupled thermal and electrical transport with decreasing film thickness due to boundary scattering. The derivation assumes that (i) the Wiedemann--Franz law is valid for the electronic contribution to thermal conductivity, (ii) Matthiessen's rule describes the thickness-dependent electron mean free path, and (iii) interface and adhesion-layer effects are neglected. Because the Lorenz number may deviate from the Sommerfeld value when phonons contribute appreciably to thermal transport, we independently measure the thermal conductivity and electrical conductivity using steady-state thermoreflectance (SSTR) and four-point probe measurements, respectively, to accurately evaluate the transport performance of the metallic thin films.

\subsection{Modeling the Effect of the Ta Adhesion/Barrier Layer}

To account for the practical implementation of Cu interconnects, we incorporate a 3~nm Ta adhesion/barrier layer into the transport analysis. The Cu/Ta interconnect is modeled as two layers connected in series for both electrical and thermal transport. The electrical resistance of each layer is given by

\begin{equation}
R_i=\frac{\rho_iL}{A_i},
\end{equation}

where $\rho_i$, $L$, and $A_i$ are the electrical resistivity, conductor length, and cross-sectional area of layer $i$, respectively. The total electrical resistance of the Cu/Ta stack is therefore

\begin{equation}
R_{\mathrm{elec}}=R_{\mathrm{Cu}}+R_{\mathrm{Ta}}.
\end{equation}

The electrical resistivity of the 3~nm Ta layer is taken as
$\rho_{\mathrm{Ta}}=2.69\times10^{-6}$~$\Omega\cdot$m,
corresponding to an electrical conductivity of
$\sigma_{\mathrm{Ta}}=3.69\times10^{5}$~S\,m$^{-1}$.
The electronic thermal conductivity of Ta is estimated using the Wiedemann--Franz relation,

\begin{equation}
\kappa_{\mathrm{Ta}}=L_0\sigma_{\mathrm{Ta}}T,
\end{equation}

where $L_0=2.44\times10^{-8}$~W$\Omega$K$^{-2}$ is the Sommerfeld Lorenz number and $T=298$~K, yielding

\[
\kappa_{\mathrm{Ta}}\approx2.66~\mathrm{W\,m^{-1}K^{-1}}.
\]

Similarly, the thermal resistance of each layer is

\begin{equation}
R_{\mathrm{th},i}=\frac{L}{\kappa_iA_i},
\end{equation}

and the total thermal resistance of the Cu/Ta stack is

\begin{equation}
R_{\mathrm{th}}=R_{\mathrm{th,Cu}}+R_{\mathrm{th,Ta}}.
\end{equation}

The corresponding effective electrical and thermal conductivities of the Cu/Ta stack are subsequently used to evaluate the transport figure of merit discussed in the main text. The electrical resistivity and thermal conductivity of Cu are taken from our previous study~\cite{islam2024evaluating}.

\section{TDTR Measurements, Sensitivity Analysis, and Uncertainty}

Time-domain thermoreflectance (TDTR) measurements are performed to determine the thermophysical parameters required for the steady-state thermoreflectance (SSTR) thermal model. The experimental configuration has been described in detail elsewhere~\cite{islam2026unveiling,islam2024evaluating}. Briefly, a Ti:sapphire laser (Spectra Physics Tsunami, $\sim$808~nm, 80~MHz) generates the pump and probe beams. The pump beam is modulated using an electro-optic modulator (EOM) to induce a periodic surface temperature oscillation, while the probe beam monitors the corresponding change in reflectivity through a balanced photodetector and lock-in amplifier. The effective pump--probe beam radius is determined using an Al-coated fused silica calibration sample (Corning HPFS 7980, $\kappa = 1.36$~W\,m$^{-1}$\,K$^{-1}$). Fitting the known thermal conductivity with the TDTR thermal model yields an effective beam radius of approximately 2.15~$\mu$m at the 20$\times$ objective. Assuming identical pump and probe beam sizes, we further verify the beam radius by measuring the thermal conductivity of a sapphire reference sample. To evaluate the applicability of TDTR for the present metal films, we calculate the measurement sensitivity to the in-plane thermal conductivity of 100~nm Ir, Mo, and Co films following the methodology reported previously for Cu, Ru and W films~\cite{islam2024evaluating,islam2026unveiling}. A 20~nm Al/60~nm Ti transducer is employed to ensure optical opacity and to laterally distribute the absorbed laser energy within the transducer before it enters the metal films. The sensitivity calculations are performed at two modulation frequencies using an effective beam radius of 2.15~$\mu$m and the thermophysical parameters listed in Table~S4.

\begin{figure}[htb]
\includegraphics[width=\textwidth]{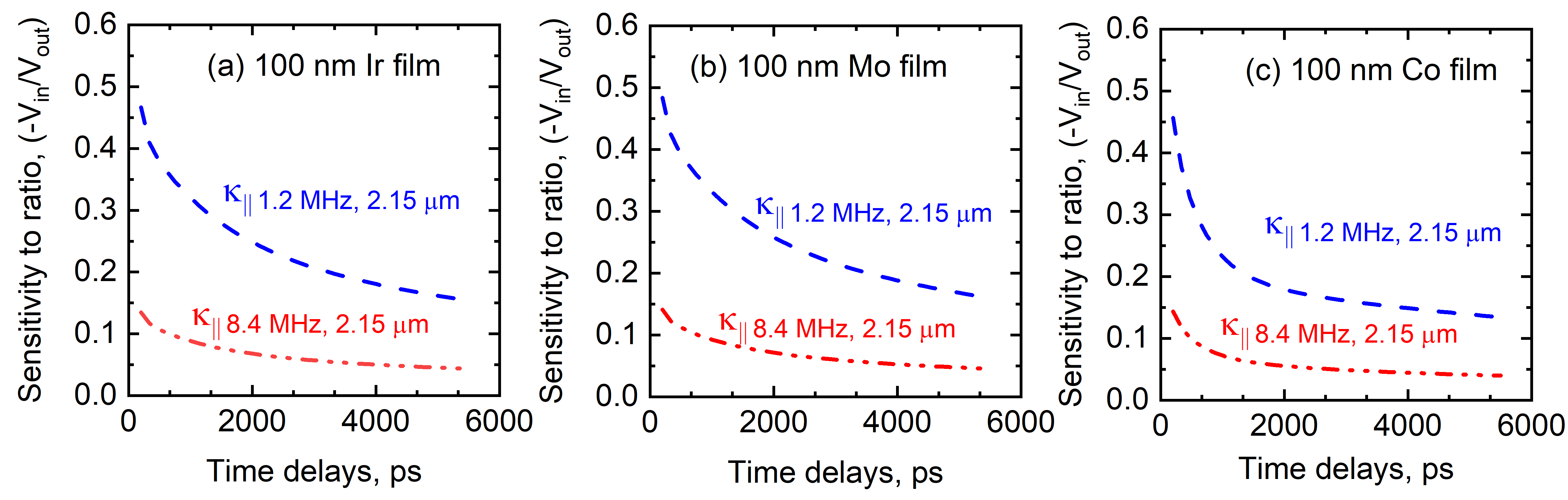}
\centering
\caption{\textbf{TDTR sensitivity for 100~nm Ir, Mo, and Co films}. TDTR sensitivity to the in-plane thermal conductivity ($\kappa_{\parallel}$) of Al/Ti-coated 100~nm Ir, Mo, and Co films for different modulation frequencies at an effective pump--probe beam radius of 2.15~$\mu$m}
\label{supp:Fig01}
\end{figure}

As shown in Fig.~\ref{supp:Fig01}, TDTR exhibits the highest sensitivity to the in-plane thermal conductivity of the 100~nm Ir, Mo, and Co films at a modulation frequency of 1.2~MHz. Nevertheless, the sensitivity decreases rapidly with decreasing film thickness, becoming insufficient for films thinner than approximately 50~nm. Under the present experimental configuration, TDTR also does not provide sufficient sensitivity to independently determine the cross-plane thermal conductivity of these metal films. Consequently, only the in-plane thermal conductivity of the 100~nm Ir, Mo, and Co films is extracted from the TDTR measurements. The sensitivity analysis further indicates that the TDTR response is more strongly influenced by the thermal conductivities of the SiO$_2$ layer and Si substrate than by those of the metal films. Therefore, the accepted literature values for SiO$_2$ and Si are fixed in the thermal model to improve the robustness of the fitting procedure~\cite{braun2019steady,hoque2021thermal,cancellieri2020interface,braun2016size,scott2018thermal}.
\setlength{\tabcolsep}{.3em}

\begin{table*}[h!]
\caption{\label{supp:tab:thermal condutvities}
Parameters used in the thermal model to determine the in-plane thermal conductivities of Ir, Mo, and Co thin films.}
\centering
\resizebox{\textwidth}{!}{
\begin{tabular}{@{}lcccc@{}}
\toprule
\textbf{Material} &
\textbf{Thickness (nm)} &
\textbf{Heat Capacity (MJ/m$^3$K)} &
\textbf{Thermal Conductivity (W/m$\cdot$K)} &
\textbf{TBR (m$^2$K/GW)} \\
\midrule

Al            & 20     & 2.42~\cite{cancellieri2020interface} & 110--120 & -- \\
Al/Ti         & --     & --                                   & --                      & 0.333~\cite{wilson2012experimental,cheaito2015thermal} \\
Ti            & 60     & 2.36~\cite{olson2021band}            & 16--19                  & -- \\

Ir            & 5--100 & 2.94~\cite{thakur2024ab}                               & 100 nm, $k_{\parallel}=16.35\pm12.4^{\mathrm{a}}$ & -- \\
Ti/Ir         & --     & --                                   & --                      & 1.33$^{\mathrm{a}}$ \\

Mo            & 5--100 & 2.56~\cite{desai1987thermodynamic}                                & 100 nm, $k_{\parallel}=28.18\pm15.6^{\mathrm{a}}$ & -- \\
Ti/Mo         & --     & --                                   & --                      & 1.33$^{\mathrm{a}}$ \\

Co            & 5--100 & 3.76 ~\cite{montague1979thermal}                               & 100 nm, $k_{\parallel}=31.77\pm15.8^{\mathrm{a}}$ & -- \\
Ti/Co         & --     & --                                   & --                      & 3.11$^{\mathrm{a}}$ \\

Metal/SiO$_2$ & --     & --                                   & --                      & 5.00~\cite{hoque2021thermal} \\
SiO$_2$
              & 100     & 1.62~\cite{braun2019steady,hoque2021thermal,braun2016size}
              & 1.45~\cite{cancellieri2020interface,braun2019steady,hoque2021thermal,braun2016size}
              & -- \\
SiO$_2$/Si    & --     & --                                   & --                      & 4.35~\cite{hoque2021thermal,braun2016size} \\
Si            & --     & 1.65~\cite{cancellieri2020interface,braun2019steady,hoque2021thermal,braun2016size}
              & 140~\cite{braun2019steady,hoque2021thermal}
              & -- \\

\bottomrule
\end{tabular}
}

\vspace{0.5em}
\begin{minipage}{\textwidth}
\scriptsize

\textbf{a.} The listed thermal conductivities correspond to the measured in-plane thermal conductivities of the 100 nm reference films determined by TDTR:  The corresponding TDTR-measured thermal boundary resistances are used in the thermal model. We propagate a 50\% uncertainty in both measured and assumed TBR values.

\textbf{b.}We perturb the thermal conductivities of Al, Ti, and Si by 5\% and that of SiO$_2$ by 3\%. Since the cross-plane thermal conductivity of the metal films cannot be resolved from the TDTR measurements, we assume it is equal to the corresponding electronic thermal conductivity and assign a 50\% uncertainty to this parameter. We perturb the film thicknesses by 3\%.

\end{minipage}
\end{table*}

\begin{figure}[htb]
\includegraphics[width=\textwidth]{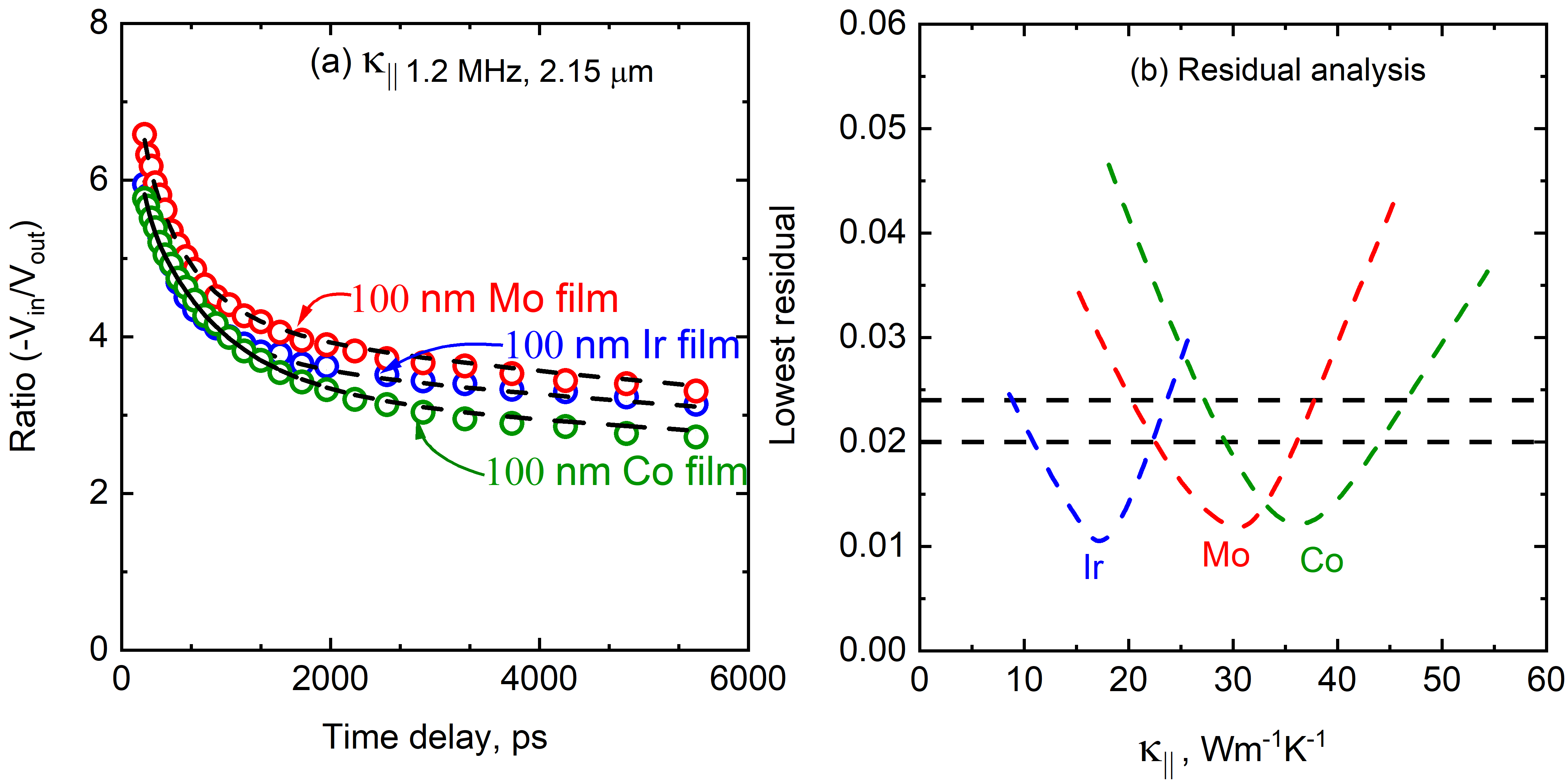}
\centering
\caption{\textbf{Theoretical fit and error analysis}.{(a)} Measured TDTR signals (symbols) and best-fit thermal model (dashed lines) for 100~nm Ir, Mo, and Co films obtained at 1.2~MHz with an effective pump--probe beam radius of 2.15~$\mu$m. \textbf{(b)} Residual error as a function of the fitted in-plane thermal conductivity ($\kappa_{\parallel}$). The dashed horizontal lines denote the 1.5\% residual criterion adopted to estimate the uncertainty in the extracted thermal conductivity.}
\label{supp:Fig02}
\end{figure}

The in-plane thermal conductivities of the Ir, Mo, and Co films are extracted by fitting the TDTR response with a five-layer heat diffusion model~\cite{cahill2004analysis, hopkins2010criteria}. The thermophysical properties, layer thicknesses, and thermal boundary resistances used in the calculations are summarized in Table~\ref{supp:tab:thermal condutvities}. Because the present TDTR configuration does not provide sufficient sensitivity to independently determine the cross-plane thermal conductivity of the Ir, Mo and Co films, we assume it is equal to the corresponding electronic thermal conductivity and use this value as a  model parameter during the fitting procedure. The experimental TDTR data are analyzed by comparing the measured ratio of the in-phase and out-of-phase voltages, $-V_{\mathrm{in}}/V_{\mathrm{out}}$, with the predictions of the thermal model. Representative fits for the 100~nm Ir, Mo, and Co films are presented in Fig.~\ref{supp:Fig02}a, demonstrating excellent agreement between the measured response and the model. The fitting procedure simultaneously determines the in-plane thermal conductivity and the thermal boundary resistance between the transducer and the metal film. The extracted values are listed in Table~\ref{supp:tab:thermal condutvities}.

To evaluate the uniqueness of the fitted parameters, we calculate residual error  by varying the in-plane thermal conductivity about their optimum values. As shown in Fig.~\ref{supp:Fig02}b, multiple parameter combinations produce comparable agreement with the experimental data. We therefore define the uncertainty associated with the fitting procedure using a residual threshold of 1.5\%, corresponding to the range of parameter pairs that reproduce the measured TDTR response within this criterion. The uncertainty in the extracted in-plane thermal conductivity includes contributions from the fitting residual, spatial variation among repeated measurements, and uncertainties in the thermal model parameters. Specifically, we propagate uncertainties associated with the thermal conductivities of Al, Ti, SiO$_2$, and Si, the thermal boundary resistances, and the assumed cross-plane thermal conductivity of the metal films. The total uncertainty, $\varepsilon_i$, is calculated as

\begin{equation}
\varepsilon_i=\sqrt{\sigma_i^2+\sum_i\Delta_i^2},
\label{supp:eq:S06}
\end{equation}

where $\sigma_i$ denotes the standard deviation obtained from measurements at multiple locations on each sample and $\Delta_i$ represents the uncertainty associated with each parameter in the thermal model~\cite{braun2019steady,feser2012probing}. The resulting uncertainties, together with the extracted in-plane thermal conductivities and thermal boundary resistances, are reported in Table~\ref{supp:tab:thermal condutvities}.

\section{SSTR Measurements, Sensitivity Analysis, and Uncertainty}

To establish the capability of steady-state thermoreflectance (SSTR) to determine the in-plane thermal conductivity of Ir, Mo, and Co films, we calculate the measurement sensitivity to the thermophysical parameters of the multilayer structure as a function of film thickness. The sensitivity analysis follows the methodology reported in our previous work~\cite{islam2024evaluating,islam2026unveiling} and the formalism developed by Braun \textit{et al.} and Yang \textit{et al.}~\cite{braun2019steady,yang2013thermal}.

\begin{figure}[htb]
\includegraphics[width=\textwidth]{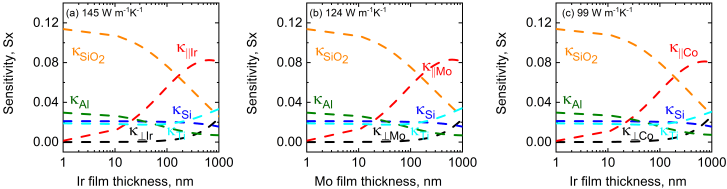}
\centering
\caption{\textbf{Sensitivity analysis of SSTR measurements.} Calculated sensitivities of the SSTR signal to the thermophysical parameters of (a) Ir, (b) Mo, and (c) Co films as a function of film thickness. The calculations are performed using the corresponding thermal model parameters listed in Table~\ref{supp:tab:thermal condutvities}. For all three materials, the sensitivity to the in-plane thermal conductivity ($\kappa_{\parallel}$) increases with increasing film thickness and exceeds the sensitivity to the cross-plane thermal conductivity ($\kappa_{\perp}$), enabling reliable extraction of the in-plane thermal conductivity over the investigated thickness range.}
\label{supp:Fig03}
\end{figure}

The sensitivity to a parameter, $x$, is defined as

\begin{equation}
S_x=\frac{\Delta T_{1.1x}-\Delta T_{0.9x}}{\Delta T_x},
\label{supp:eq:S09}
\end{equation}

where $\Delta T_x$ is the temperature rise predicted by the steady-state heat diffusion model using the nominal material properties, and $\Delta T_{1.1x}$ and $\Delta T_{0.9x}$ correspond to 10\% positive and negative perturbations of the parameter, respectively. The calculations employ the thermophysical properties listed in Table~\ref{supp:tab:thermal condutvities}. Figure~\ref{supp:Fig03} shows the calculated sensitivities for Ir, Mo, and Co films using an effective pump--probe beam radius of approximately 2.15~$\mu$m. For all three materials, the SSTR response is considerably more sensitive to the in-plane thermal conductivity ($\kappa_{\parallel}$) than to the cross-plane thermal conductivity ($\kappa_{\perp}$). This behavior arises because the thermal conductivity of the metal films greatly exceeds that of the underlying SiO$_2$ layer, promoting lateral heat transport within the film and producing a stronger in-plane temperature gradient~\cite{braun2019steady,hoque2021high}. Consequently, SSTR primarily probes the in-plane thermal conductivity of the metallic films. Figure~\ref{supp:Fig03} shows that SSTR signal remains more sensitive to the SiO$_2$ thermal conductivity than to the metal-film thermal conductivity over a broad thickness range. We use the accepted literature value of $\kappa_{\mathrm{SiO_2}}=1.45$~W\,m$^{-1}$\,K$^{-1}$ range~\cite{braun2016size,braun2019steady,cancellieri2020interface,hoque2021thermal}. Likewise, literature values are adopted for the thermal conductivity of Si and for the heat capacities and thermal boundary resistances used in the thermal model (Table~\ref{supp:tab:thermal condutvities})~\cite{braun2016size,braun2019steady,cancellieri2020interface,hoque2021thermal}. These results demonstrate that the chosen SSTR configuration provides sufficient sensitivity to accurately determine the in-plane thermal conductivity of Ir, Mo, and Co thin films.

\begin{figure}[htbp]
\centering
\includegraphics[
    width=\linewidth,
    height=0.82\textheight,
    keepaspectratio
]{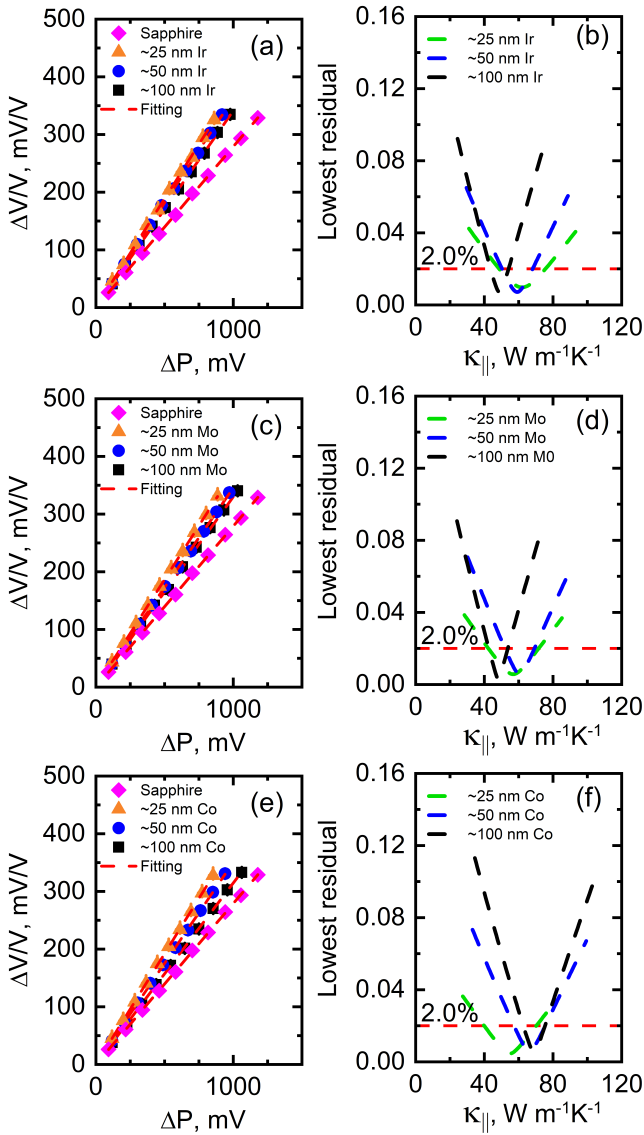}
\caption{\textbf{Representative SSTR fitting and residual analysis.} Experimental SSTR data (symbols) and best-fit thermal model (dashed lines) for (a) Ir, (c) Mo, and (e) Co films with thicknesses of approximately 25, 50, and 100~nm, together with a sapphire reference sample used for gamma calibration. The corresponding residual error as a function of the fitted in-plane thermal conductivity ($\kappa_{\parallel}$) is shown in (b), (d), and (f). The dashed horizontal line indicates the 2.0\% residual threshold adopted to determine the uncertainty in the extracted thermal conductivity.}
\label{supp:Fig04}
\end{figure}

The in-plane thermal conductivities of the Ir, Mo, and Co films are determined by fitting the SSTR measurements with a five-layer heat diffusion model consisting of the Al/Ti transducer, metal film, SiO$_2$, and Si substrate~\cite{braun2019steady,hoque2021thermal,islam2024evaluating}. Material properties not directly determined by SSTR, including the thermal boundary resistances are obtained from TDTR measurements or established literature values, as summarized in Table~\ref{supp:tab:thermal condutvities}. Figures~\ref{supp:Fig04}(a), (c), and (e) compare the measured normalized thermoreflectance signal ($\Delta V/V$) with the corresponding thermal-model predictions as a function of the pump-induced voltage change ($\Delta P$). A sapphire reference sample with well-established thermophysical properties is used to determine the calibration factor ($\gamma$), which is subsequently fixed for all measurements. The only adjustable parameter in the fitting procedure is the in-plane thermal conductivity, $\kappa_{\parallel}$. The fitting robustness is evaluated through the residual analysis shown in Figs.~\ref{supp:Fig04}(b), (d), and (f), where the residual error is plotted as a function of the assumed in-plane thermal conductivity. The minimum of each curve identifies the best-fit value of $\kappa_{\parallel}$, while the dashed horizontal line denotes the 2.0\% residual threshold adopted to define the acceptable fitting window. This criterion is selected based on the maximum deviation between the experimental data and the thermal model and provides a consistent estimate of the uncertainty associated with the fitting procedure. The total uncertainty reported in the thermal conductivity is calculated using Eq.~\ref{supp:eq:S06}, combining the fitting residual, spot-to-spot measurement variability, and propagated uncertainties in the thermophysical properties and thermal boundary resistances listed in Table~\ref{supp:tab:thermal condutvities}.

\begin{figure}[h!]
\includegraphics[width=\textwidth]{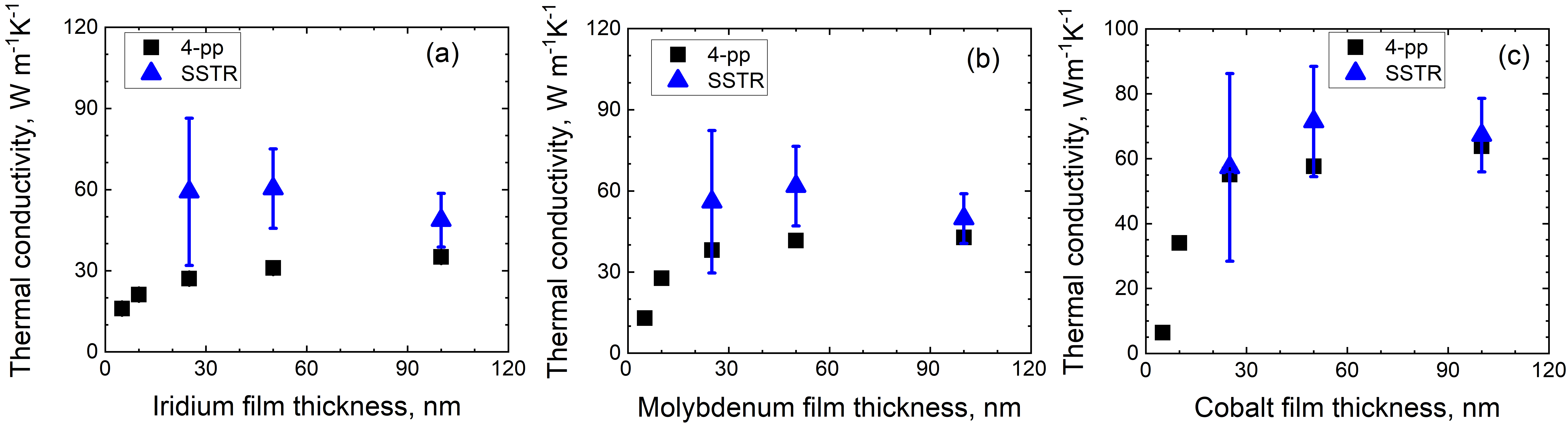}
\centering
\caption{\textbf{Comparison of electronic and total thermal conductivities of Ir, Mo, and Co films.} (a)--(c) In-plane thermal conductivities of Ir, Mo, and Co films measured by SSTR (blue triangles) together with the electronic thermal conductivities estimated from four-point probe resistivity measurements using the Wiedemann--Franz law (black squares). The measured thermal conductivity is consistently higher than the electronic contribution, consistent with previously reported phonon-mediated heat transport in these metals~\cite{wen2020large,perez2022dominant,tong2019comprehensive}.}
\label{supp:Fig05}
\end{figure}

 Figure~\ref{supp:Fig05} compares the electronic thermal conductivity with the total in-plane thermal conductivity measured by SSTR. The electronic thermal conductivity is independently estimated from the measured electrical resistivity using the Wiedemann--Franz law with the Sommerfeld Lorenz number. For all three metals, the SSTR-measured thermal conductivity consistently exceeds the electronic contribution estimated from the Wiedemann--Franz relation. This observation is consistent with previous studies reporting a finite phonon contribution to thermal transport in Ir, Mo, and Co.\cite{wen2020large,perez2022dominant,tong2019comprehensive} A quantitative separation of the electronic and phononic contributions is beyond the scope of the present work. The larger uncertainties observed for films thinner than approximately 50~nm arise from the reduced sensitivity of the SSTR measurements to the in-plane thermal conductivity in this thickness regime.

\bibliographystyle{unsrt}
\bibliography{Refs}

@article{gall2016electron,
  title={Electron mean free path in elemental metals},
  author={Gall, Daniel},
  journal={Journal of applied physics},
  volume={119},
  number={8},
  pages={085101},
  year={2016},
  publisher={AIP Publishing LLC}
}

@article{braun2019steady,
  title={A steady-state thermoreflectance method to measure thermal conductivity},
  author={Braun, Jeffrey L and Olson, David H and Gaskins, John T and Hopkins, Patrick E},
  journal={Review of Scientific Instruments},
  volume={90},
  number={2},
  year={2019},
  publisher={AIP Publishing}
}

@article{hoque2021high,
  title={High in-plane thermal conductivity of aluminum nitride thin films},
  author={Hoque, Md Shafkat Bin and Koh, Yee Rui and Braun, Jeffrey L and Mamun, Abdullah and Liu, Zeyu and Huynh, Kenny and Liao, Michael E and Hussain, Kamal and Cheng, Zhe and Hoglund, Eric R and others},
  journal={ACS nano},
  volume={15},
  number={6},
  pages={9588--9599},
  year={2021},
  publisher={ACS Publications}
}

@article{hoque2021thermal,
  title={Thermal conductivity measurements of sub-surface buried substrates by steady-state thermoreflectance},
  author={Hoque, Md Shafkat Bin and Koh, Yee Rui and Aryana, Kiumars and Hoglund, Eric R and Braun, Jeffrey L and Olson, David H and Gaskins, John T and Ahmad, Habib and Elahi, Mirza Mohammad Mahbube and Hite, Jennifer K and others},
  journal={Review of Scientific Instruments},
  volume={92},
  number={6},
  year={2021},
  publisher={AIP Publishing}
}

@article{lo2019enhancing,
  title={Enhancing interconnect reliability and performance by converting tantalum to 2D layered tantalum sulfide at low temperature},
  author={Lo, Chun-Li and Catalano, Massimo and Khosravi, Ava and Ge, Wanying and Ji, Yujin and Zemlyanov, Dmitry Y and Wang, Luhua and Addou, Rafik and Liu, Yuanyue and Wallace, Robert M and others},
  journal={Advanced Materials},
  volume={31},
  number={30},
  pages={1902397},
  year={2019},
  publisher={Wiley Online Library}
}

@article{zhan2020effect,
  title={Effect of Thermal Boundary Resistance between the Interconnect Metal and Dielectric Interlayer on Temperature Increase of Interconnects in Deeply Scaled VLSI},
  author={Zhan, Tianzhuo and Oda, Kaito and Ma, Shuaizhe and Tomita, Motohiro and Jin, Zhicheng and Takezawa, Hiroki and Mesaki, Kohei and Wu, Yen-Ju and Xu, Yibin and Matsukawa, Takashi and others},
  journal={ACS applied materials \& interfaces},
  volume={12},
  number={19},
  pages={22347--22356},
  year={2020},
  publisher={ACS Publications}
}

@article{cancellieri2020interface,
  title={Interface and layer periodicity effects on the thermal conductivity of copper-based nanomultilayers with tungsten, tantalum, and tantalum nitride diffusion barriers},
  author={Cancellieri, Claudia and Scott, Ethan A and Braun, Jeffrey and King, Sean W and Oviedo, Ron and Jezewski, Christopher and Richards, John and La Mattina, Fabio and Jeurgens, Lars PH and Hopkins, Patrick E},
  journal={Journal of Applied Physics},
  volume={128},
  number={19},
  year={2020},
  publisher={AIP Publishing}
}

@article{tong2019comprehensive,
  title={Comprehensive first-principles analysis of phonon thermal conductivity and electron-phonon coupling in different metals},
  author={Tong, Zhen and Li, Shouhang and Ruan, Xiulin and Bao, Hua},
  journal={Physical review B},
  volume={100},
  number={14},
  pages={144306},
  year={2019},
  publisher={APS}
}

@article{gall2020search,
  title={The search for the most conductive metal for narrow interconnect lines},
  author={Gall, Daniel},
  journal={Journal of Applied Physics},
  volume={127},
  number={5},
  year={2020},
  publisher={AIP Publishing}
}

@article{islam2024evaluating,
  title={Evaluating size effects on the thermal conductivity and electron-phonon scattering rates of copper thin films for experimental validation of Matthiessen’s rule},
  author={Islam, Md Rafiqul and Karna, Pravin and Tomko, John A and Hoglund, Eric R and Hirt, Daniel M and Hoque, Md Shafkat Bin and Zare, Saman and Aryana, Kiumars and Pfeifer, Thomas W and Jezewski, Christopher and others},
  journal={Nature Communications},
  volume={15},
  number={1},
  pages={9167},
  year={2024},
  publisher={Nature Publishing Group UK London}
}

@article{feser2012probing,
  title={Probing anisotropic heat transport using time-domain thermoreflectance with offset laser spots},
  author={Feser, Joseph P and Cahill, David G},
  journal={Review of Scientific Instruments},
  volume={83},
  number={10},
  year={2012},
  publisher={AIP Publishing}
}

@article{perez2022dominant,
  title={Dominant energy carrier transitions and thermal anisotropy in epitaxial iridium thin films},
  author={Perez, Christopher and Jog, Atharv and Kwon, Heungdong and Gall, Daniel and Asheghi, Mehdi and Kumar, Suhas and Park, Woosung and Goodson, Kenneth E},
  journal={Advanced Functional Materials},
  volume={32},
  number={45},
  pages={2207781},
  year={2022},
  publisher={Wiley Online Library}
}

@article{islam2026unveiling,
  title={Unveiling phonon contributions to thermal conductivity and the applicability of the Wiedemann—Franz law in ruthenium and tungsten thin films},
  author={Islam, Md Rafiqul and Karna, Pravin and Bhatt, Niraj and Thakur, Sandip and Heinrich, Helge and Hirt, Daniel M and Zare, Saman and Jezewski, Christopher and Lee, Rinus TP and Tapily, Kandabara and others},
  journal={Advanced Functional Materials},
  volume={36},
  number={12},
  pages={e11592},
  year={2026},
  publisher={Wiley Online Library}
}

@article{wen2020large,
  title={Large lattice thermal conductivity, interplay between phonon-phonon, phonon-electron, and phonon-isotope scatterings, and electrical transport in molybdenum from first principles},
  author={Wen, Shihao and Ma, Jinlong and Kundu, Ashis and Li, Wu},
  journal={Physical Review B},
  volume={102},
  number={6},
  pages={064303},
  year={2020},
  publisher={APS}
}

@article{cahill2004analysis,
  title={Analysis of heat flow in layered structures for time-domain thermoreflectance},
  author={Cahill, David G and others},
  journal={Review of scientific instruments},
  volume={75},
  number={12},
  pages={5119},
  year={2004}
}

@article{hopkins2010criteria,
  title={Criteria for cross-plane dominated thermal transport in multilayer thin film systems during modulated laser heating},
  author={Hopkins, Patrick E and Serrano, Justin R and Phinney, Leslie M and Kearney, Sean P and Grasser, Thomas W and Harris, C Thomas},
  year={2010}
}

@article{desai1987thermodynamic,
  title={Thermodynamic properties of manganese and molybdenum},
  author={Desai, Pramond D},
  journal={Journal of physical and chemical reference data},
  volume={16},
  number={1},
  pages={91--108},
  year={1987},
  publisher={American Institute of Physics for the National Institute of Standards and~…}
}

@article{olson2021band,
  title={Band alignment and defects influence the electron--phonon heat transport mechanisms across metal interfaces},
  author={Olson, David H and Sales, Maria G and Tomko, John A and Lu, Teng-Fei and Prezhdo, Oleg V and McDonnell, Stephen J and Hopkins, Patrick E},
  journal={Applied Physics Letters},
  volume={118},
  number={16},
  year={2021},
  publisher={AIP Publishing}
}

@article{wilson2012experimental,
  title={Experimental validation of the interfacial form of the Wiedemann-Franz law},
  author={Wilson, RB and Cahill, David G},
  journal={Physical review letters},
  volume={108},
  number={25},
  pages={255901},
  year={2012},
  publisher={APS}
}

@article{cheaito2015thermal,
  title={Thermal flux limited electron Kapitza conductance in copper-niobium multilayers},
  author={Cheaito, Ramez and Hattar, Khalid and Gaskins, John T and Yadav, Ajay K and Duda, John C and Beechem, Thomas E and Ihlefeld, Jon F and Piekos, Edward S and Baldwin, Jon K and Misra, Amit and others},
  journal={Applied Physics Letters},
  volume={106},
  number={9},
  year={2015},
  publisher={AIP Publishing}
}

@article{braun2016size,
  title={Size effects on the thermal conductivity of amorphous silicon thin films},
  author={Braun, Jeffrey L and Baker, Christopher H and Giri, Ashutosh and Elahi, Mirza and Artyushkova, Kateryna and Beechem, Thomas E and Norris, Pamela M and Leseman, Zayd C and Gaskins, John T and Hopkins, Patrick E},
  journal={Physical Review B},
  volume={93},
  number={14},
  pages={140201},
  year={2016},
  publisher={APS}
}

@inproceedings{lesniewska2020dielectric,
  title={Dielectric reliability study of 21 nm pitch interconnects with barrierless Ru fill},
  author={Le{\'s}niewska, A and Roussel, Philippe J and Tierno, Davide and Gonzalez, V Vega and van der Veen, Marleen H and Verdonck, Patrick and Jourdan, Nicolas and Wilson, Christopher J and T{\H{o}}kei, Zs and Croes, Kris},
  booktitle={2020 IEEE international reliability physics symposium (IRPS)},
  pages={1--6},
  year={2020},
  organization={IEEE}
}

@inproceedings{lesniewska2021reliability,
  title={Reliability of a DME Ru Semidamascene scheme with 16 nm wide Airgaps},
  author={Le{\'s}niewska, A and Pedreira, O Varela and Lofrano, Melina and Murdoch, Gayle and van der Veen, Marleen and Dangol, Anish and Horiguchi, Naoto and T{\H{o}}kei, Zs and Croes, Kris},
  booktitle={2021 IEEE International Reliability Physics Symposium (IRPS)},
  pages={1--6},
  year={2021},
  organization={IEEE}
}

@inproceedings{griggio2018reliability,
  title={Reliability of dual-damascene local interconnects featuring cobalt on 10 nm logic technology},
  author={Griggio, F and Palmer, James and Pan, F and Toledo, N and Schmitz, Anthony and Tsameret, Ilan and Kasim, R and Leatherman, G and Hicks, Jeffery and Madhavan, A and others},
  booktitle={2018 IEEE International Reliability Physics Symposium (IRPS)},
  pages={6E--3},
  year={2018},
  organization={IEEE}
}

@article{black1969electromigration,
  title={Electromigration—A brief survey and some recent results},
  author={Black, James R},
  journal={IEEE Transactions on Electron Devices},
  volume={16},
  number={4},
  pages={338--347},
  year={1969},
  publisher={IEEE}
}

@inproceedings{edelstein1997full,
  title={Full copper wiring in a sub-0.25/spl mu/m CMOS ULSI technology},
  author={Edelstein, Dan and Heidenreich, J and Goldblatt, R and Cote, W and Uzoh, C and Lustig, N and Roper, P and McDevitt, T and Motsiff, W and Simon, A and others},
  booktitle={International Electron Devices Meeting. IEDM Technical Digest},
  pages={773--776},
  year={1997},
  organization={IEEE}
}

@inproceedings{zahedmanesh2019copper,
  title={Copper electromigration; prediction of scaling limits},
  author={Zahedmanesh, Houman and Pedreira, O Varela and Wilson, Chris and T{\H{o}}kei, Zsolt and Croes, Kristof},
  booktitle={Proceedings of IEEE International Interconnect Technology Conference (IITC)},
  pages={3--5},
  year={2019},
  organization={IEEE}
}

@article{soulie2024selecting,
  title={Selecting alternative metals for advanced interconnects},
  author={Souli{\'e}, Jean-Philippe and Sankaran, Kiroubanand and Van Troeye, Benoit and Le{\'s}niewska, Alicja and Varela Pedreira, Olalla and Oprins, Herman and Delie, Gilles and Fleischmann, Claudia and Boakes, Lizzie and Rolin, C{\'e}dric and others},
  journal={Journal of Applied Physics},
  volume={136},
  number={17},
  year={2024},
  publisher={AIP Publishing}
}

@book{haynes2016crc,
  title={CRC handbook of chemistry and physics},
  author={Haynes, William M},
  year={2016},
  publisher={CRC press}
}

@article{krishnan2007towards,
  title={Towards a thermal Moore's law},
  author={Krishnan, Shankar and Garimella, Suresh V and Chrysler, Gregory M and Mahajan, Ravi V},
  journal={IEEE Transactions on advanced packaging},
  volume={30},
  number={3},
  pages={462--474},
  year={2007},
  publisher={IEEE}
}

@article{seshan2012scaling,
  title={Scaling—Its Effects on Heat Generation and Cooling of Devices. A “Thermal Moore’s” Law?},
  author={Seshan, Krishna},
  journal={Handbook of Thin Film Deposition},
  pages={41},
  year={2012},
  publisher={William Andrew}
}

@article{boakes2024selection,
  title={Selection of alternative local interconnect metals: Beyond traditional criteria towards sustainable and secure supply chains},
  author={Boakes, Lizzie and Ragnarsson, Lars-{\AA}ke and Rolin, C{\'e}dric and Adelmann, Christoph},
  journal={arXiv preprint arXiv:2401.02864},
  year={2024}
}

@article{chavez2014reduction,
  title={Reduction of the thermal conductivity in free-standing silicon nano-membranes investigated by non-invasive Raman thermometry},
  author={Ch{\'a}vez-Angel, Emigdio and Reparaz, Juan Sebastian and Gomis-Bresco, Jordi and Wagner, Markus R and Cuffe, John and Graczykowski, Bart{\l}omiej and Shchepetov, Andrey and Jiang, Hua and Prunnila, Mika and Ahopelto, Jouni and others},
  journal={APL materials},
  volume={2},
  number={1},
  year={2014},
  publisher={AIP Publishing}
}

@inproceedings{fuchs1938conductivity,
  title={The conductivity of thin metallic films according to the electron theory of metals},
  author={Fuchs, K},
  booktitle={Mathematical Proceedings of the Cambridge Philosophical Society},
  volume={34},
  number={1},
  pages={100--108},
  year={1938},
  organization={Cambridge University Press}
}

@article{montague1979thermal,
  title={Thermal diffusivities of hafnium and cobalt from 300 to 1000 K},
  author={Montague, Skiles A and Draper, Clifton W and Rosenblatt, Gerd M},
  journal={Journal of Physics and Chemistry of Solids},
  volume={40},
  number={12},
  pages={987--992},
  year={1979},
  publisher={Elsevier}
}

@article{thakur2024ab,
  title={Ab initio thermodynamic properties of iridium: A high-pressure and high-temperature study},
  author={Thakur, Balaram and Gong, Xuejun and Dal Corso, Andrea},
  journal={Computational Materials Science},
  volume={234},
  pages={112797},
  year={2024},
  publisher={Elsevier}
}

@article{scott2018thermal,
  title={Thermal resistance and heat capacity in hafnium zirconium oxide (Hf1--xZrxO2) dielectrics and ferroelectric thin films},
  author={Scott, Ethan A and Smith, Sean W and Henry, M David and Rost, Christina M and Giri, Ashutosh and Gaskins, John T and Fields, Shelby S and Jaszewski, Samantha T and Ihlefeld, Jon F and Hopkins, Patrick E},
  journal={Applied Physics Letters},
  volume={113},
  number={19},
  year={2018},
  publisher={AIP Publishing}
}

@article{yang2013thermal,
  title={Thermal property microscopy with frequency domain thermoreflectance},
  author={Yang, Jia and Maragliano, Carlo and Schmidt, Aaron J},
  journal={Review of Scientific Instruments},
  volume={84},
  number={10},
  year={2013},
  publisher={AIP Publishing}
}

\end{document}